\documentclass[conference]{IEEEtran}
\usepackage[noadjust]{cite}
\usepackage{amsmath,amssymb,amsfonts}
\usepackage[bookmarks=false]{hyperref}
\hypersetup{draft}

\usepackage{graphicx}
\usepackage{textcomp}
\usepackage{xcolor}
\newcommand{\etal}{\textit{et al. }}
\usepackage{caption}
\usepackage{amsmath}
\usepackage{amssymb}
\usepackage{graphicx}
\usepackage{enumitem}

\usepackage{algorithmicx}
\usepackage[ruled,linesnumbered]{algorithm2e}
\usepackage{algpseudocode}
\usepackage{multirow}
\usepackage{graphicx}
\usepackage{subfigure}
\usepackage{booktabs}
\usepackage{flushend}

\begin{document}

\title{Towards Adversarial-Resilient Deep Neural Networks for False Data Injection Attack Detection in Power Grids}

\author{
    \IEEEauthorblockN{Jiangnan Li$^{\ast}$, Yingyuan Yang$^{\S}$, Jinyuan Stella Sun$^{\ast}$, Kevin Tomsovic$^{\ast}$, Hairong Qi$^{\ast}$}
    \IEEEauthorblockA{$^{\ast}$The University of Tennessee, Knoxville, jiangnanutk@gmail.com, {\{jysun, tomsovic, hqi\}@utk.edu}
    \\$^{\S}$University of Illinois Springfield, yyang260@uis.edu}
}

\IEEEoverridecommandlockouts
\IEEEpubid{\makebox[\columnwidth]{\textbf{This paper has been accepted by IEEE ICCCN 2023.} \hfill} \hspace{\columnsep}\makebox[\columnwidth]{ }}

 \maketitle

\maketitle

\begin{abstract}

False data injection attacks (FDIAs) pose a significant security threat to power system state estimation. To detect such attacks, recent studies have proposed machine learning (ML) techniques, particularly deep neural networks (DNNs). However, most of these methods fail to account for the risk posed by adversarial measurements, which can compromise the reliability of DNNs in various ML applications. In this paper, we present a DNN-based FDIA detection approach that is resilient to adversarial attacks. We first analyze several adversarial defense mechanisms used in computer vision and show their inherent limitations in FDIA detection. We then propose an adversarial-resilient DNN detection framework for FDIA that incorporates random input padding in both the training and inference phases. Our simulations, based on an IEEE standard power system, demonstrate that this framework significantly reduces the effectiveness of adversarial attacks while having a negligible impact on the DNNs' detection performance.

\end{abstract}


\section{Introduction}

State estimation is a critical application in power grids. The state estimator in the control center uses analog measurements and status data from remote sensors to determine the network topology and estimate voltage, current magnitudes, and phase. The estimator also employs a built-in residue-based bad data detection mechanism to remove bad data and filter measurement errors. The refined data is then used by operators and advanced applications to compute market prices and make operational decisions.

False data injection attacks (FDIAs), first proposed by Liu \etal in 2009 \cite{liu2009false}, are a well-known attack vector in state estimation. They allow attackers to inject malicious false data into legitimate measurements while bypassing bad data detection, causing the estimator to output incorrect system states. These attacks can disturb critical operations, such as contingency analysis, that rely on the estimated state. Several types of FDIAs have been developed, including FDIA with incomplete information \cite{rahman2012false}, blind FDIA \cite{yu2015blind}, and outage masking \cite{7433442}.

\begin{figure}[htbp]
\centerline{\includegraphics[width=0.85\linewidth]{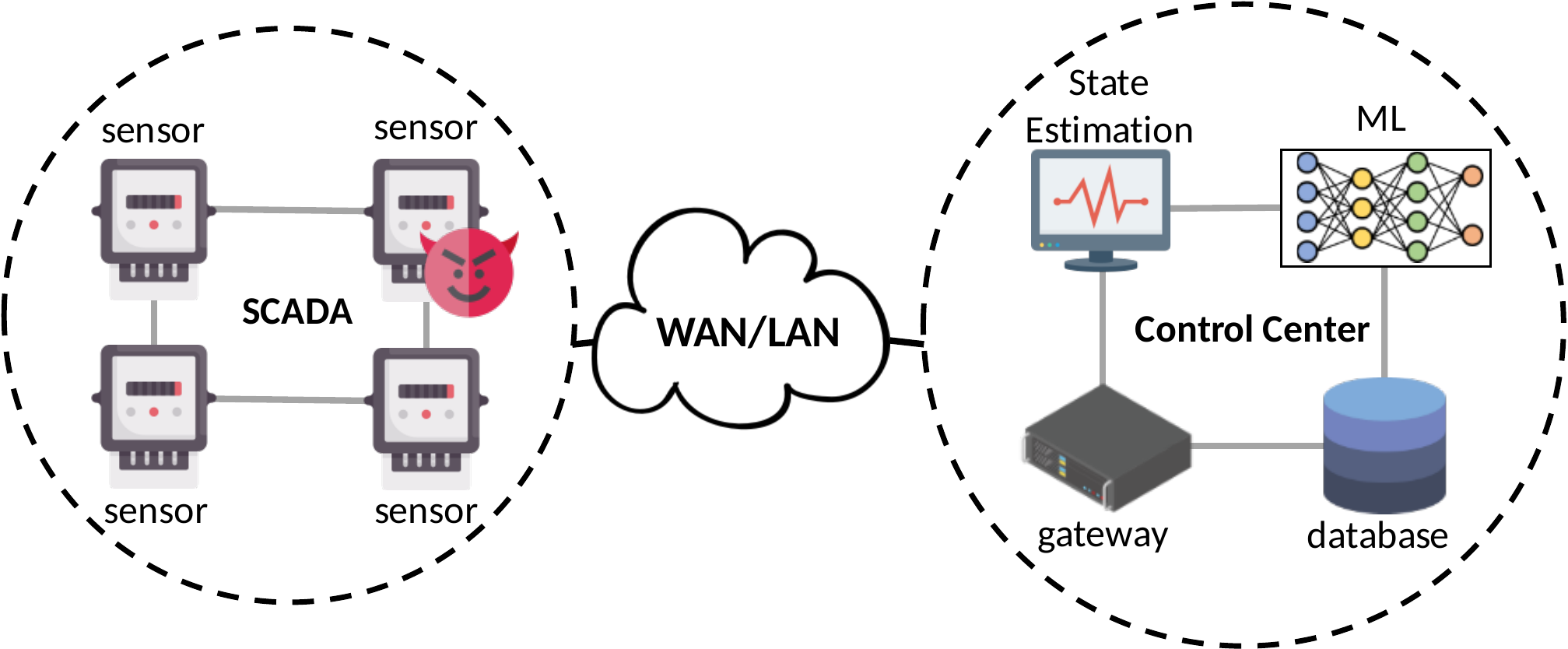}}
\caption{Illustration of DNN-based FDIA detection. The attacker compromises a subset of the sensors and inject false data. A trained DNN is placed in the control center to detect and remove malicious measurements.}
\label{exampleFDIA}
\end{figure}

To mitigate the threat posed by FDIAs, various detection methods have been proposed in the literature, such as measurement protection \cite{bi2011defending} and detection using phasor measurement units (PMUs) \cite{yang2017optimal}. In recent years, machine learning (ML), especially deep neural networks (DNNs), has become a key technique for detecting FDIAs \cite{ozay2015machine, yan2016detection, he2017real, deng2018false, james2018online, ayad2018detection, ashrafuzzaman2018detecting, niu2019dynamic, wang2020detection}. The basic architecture of DNN-based FDIA detection is shown in Fig. \ref{exampleFDIA}. DNNs leverage the statistical properties of measurement data and can achieve state-of-the-art detection performance. Furthermore, DNNs are usually pure software systems and can be deployed in the control center without requiring additional equipment or infrastructure upgrades, making them ideal defense techniques against FDIAs.

Recent research in the field of artificial intelligence (AI) has demonstrated that even well-performing deep neural networks (DNNs) are highly vulnerable to adversarial attacks. By adding well-crafted perturbations to a DNN's input, an attacker can deceive the network into outputting incorrect decisions. Such attacks have also been shown to be effective in power system applications \cite{chen2018machine, chen2019exploiting, tian2019adaptive, li2020searchfromfree}, including fault detection, isolation, and analysis (FDIA) based on DNNs \cite{li2021conaml} and state estimation \cite{liu2019adversarial}. Therefore, it is essential to consider the risks of adversarial attacks and investigate more robust DNNs that are resilient to such attacks for FDIA detection. However, typical approaches for mitigating adversarial attacks are primarily designed for computer vision applications and make specific assumptions that are not necessarily applicable to FDIA detection. For example, adversarial detection approaches often assume that adversarial examples follow a different distribution from normal inputs \cite{metzen2017detecting}. Nevertheless, such assumptions may be infeasible in FDIA detection due to physical system constraints, as demonstrated in this paper. Thus, effectively mitigating adversarial attacks in DNN-based FDIA detection represents a significant research challenge, which is the primary contribution of this paper.

In this work, we investigate defense mechanisms to mitigate adversarial attacks in DNN-based FDIA detection. As a baseline, we evaluate three typical defense methods in the computer vision domain and highlight their intrinsic limitations in preventing adversarial attacks in FDIA detection. Subsequently, we propose a general random input padding framework for DNN-based FDIA detection. Our framework can be easily adapted to different ML models used in existing research \cite{ozay2015machine, yan2016detection, he2017real, deng2018false, james2018online, ayad2018detection, ashrafuzzaman2018detecting, niu2019dynamic, wang2020detection} with simple changes to the configuration during training and detection deployment. We evaluate our approach using synthetic datasets generated from the IEEE standard 118-bus system. Our results demonstrate that the proposed framework can significantly mitigate adversarial attacks. Our main contributions are as follows:

\begin{itemize}

\item We highlight the need for robust DNN-based methods that are resilient to adversarial attacks and summarize the adversarial defense properties and requirements in FDIA detection.

\item We demonstrate that typical adversarial defense approaches (defensive distillation, adversarial training, and adversarial detection) in the computer vision domain are insufficient in mitigating adversarial attacks in FDIA detection, as shown through theoretical analysis and simulations.

\item We propose a general defense framework that randomly pads the model input during both the training and testing stages. The framework reduces the effectiveness of adversarial perturbations and can be seamlessly adapted to DNNs proposed in the literature.

\item We conduct simulations based on the standard IEEE 118-bus power system. The results show that our framework can effectively mitigate adversarial attacks while inducing minimal degradation in the models' detection accuracy of normal inputs.

\end{itemize}

The rest of the paper is organized as follows. The related work is presented in Section \ref{sec:related}.
Section \ref{sec:attack} presents the adversarial attacks in DNN-based FDIA detection. We  discuss the limitations of the existing defense approaches and propose our framework in Section \ref{sec:defense}. Simulation results are presented in Section \ref{sec:implementation}. Section \ref{sec:discussion} discusses the limitations and future work. Finally, Section \ref{sec:conclusion} concludes the paper.

\section{Related Work} \label{sec:related}

\subsection{FDIA Detection}

Since FDIA is considered as a serious threat to power system security, many studies and countermeasures are proposed. \cite{jiongcong2016impact} analyzed the impact of FDIA with static security assessment and claimed that FDIA can deceive the operator to make wrong actions, such as load shedding. Khalaf \etal studied and evaluated FDIA under wide area protection settings and showed that the FDIA can affect the grids' operation and stability \cite{8585954}. Bi and Zhang utilized graph theory to locate the meters to be protected \cite{bi2014graphical}. Some studies took advantage of new hardware devices, such as PMUs. The PMUs' measurements are synchronized to GPS signals and will increase the barrier for FDIA attacks. \cite{kim2011strategic} and \cite{chen2006placement} investigated the placement strategies of PMUs. 

Detection that utilized ML techniques became popular in the literature. In 2016, Ozay \etal first utilized ML techniques to detect FDIA \cite{ozay2015machine}. They evaluated the detection performance of traditional ML algorithms, such as K-Nearest Neighbour (KNN), support vector machine (SVM), and Sparse Logistic Regression (SLR), and demonstrated that ML achieved convincing detection accuracy with fine-tuned parameters. After that, Yan \etal studied the performance of a supervised learning classifier on detecting both direct and stealth FDIA \cite{yan2016detection}. He \etal employed deep learning (DL) for FDIA and electricity theft detection \cite{he2017real}.  In 2018, \cite{ashrafuzzaman2018detecting} evaluated the performance of plain DNN in FDIA attacks. Thereafter, different DNNs were designed. The studies in \cite{deng2018false}\cite{james2018online}\cite{ayad2018detection} used recurrent neural networks (RNNs) for FDIA detection. They evaluated the DNNs with different scale power systems and all achieves high detection results. Niu \etal designed a DNN that adopted both convolutional neural network (CNN) and RNN to detect FDIA in dynamic time-series measurement data \cite{niu2019dynamic}. False data usually follows a different distribution from legitimate data. Wang \etal trained an auto-encoder with pure normal measurement data and employed the reconstruction loss as the metric for FDIA detection \cite{wang2020detection}. 

\subsection{Adversarial Attacks}

Recent research in the AI domain has demonstrated that DNNs are vulnerable to adversarial attacks. In 2013, adversarial examples to DNNs were discovered by Szegedy \etal \cite{szegedy2013intriguing}. By adding a small crafted perturbation to the legitimate inputs, the DNNs will be led to output wrong results. After that, different adversarial attack algorithms were proposed. Goodfellow \etal presented the Fast Gradient Sign Method (FGSM) that utilized the signed gradient values to generate perturbations \cite{goodfellow2014explaining}. After that, the Fast Gradient Method by Rozsa \etal used the gradient values directly \cite{rozsa2016adversarial}. Other well-known adversarial attack algorithms include the DeepFool \cite{moosavi2016deepfool} and iterative attack \cite{kurakin2016adversarial}. 

The potential threat of adversarial attacks to critical infrastructures also draws attention in recent research. In 2018, Chen \etal investigated the effect of adversarial examples with both categorical and sequential applications in power systems \cite{chen2018machine}. They then designed adversarial attacks for regression models used for load forecasting \cite{chen2019exploiting}. Tian \etal extended \cite{chen2018machine} and proposed an adaptive normalized attack for power system ML applications \cite{tian2019adaptive}. \cite{li2021conaml} proposed constrained adversarial machine learning in CPS applications, and demonstrated an attacker could generate adversarial examples that met the intrinsic constraints defined by physical systems, such as the residual-based detection in state estimation.

The defense methods against adversarial attacks draw attention in both security and AI communities. The typical defense approaches include model distillation \cite{papernot2016distillation}, adversarial detection \cite{ma2019nic}\cite{metzen2017detecting}\cite{xu2017feature}, adversarial training \cite{kurakin2016adversarial}\cite{shafahi2019adversarial}\cite{tramer2017ensemble}, input reconstruction \cite{gu2014towards}\cite{meng2017magnet}, stochastic methods \cite{meng2017magnet}\cite{xie2018mitigating}, and so on. 
\section{Adversarial Attack in FDIA Detection} \label{sec:attack}

\subsection{Background: False Data Injection Attacks (FDIA)}

Power systems utilize state estimation to estimate the state of each bus by analyzing the other sensor's measurements. FDIA enables an attacker to generate a false measurement vector $\textbf{a}$ to be added to legitimate $\textbf{z}$, so that the polluted measurements will be $\textbf{z}_{a} = \textbf{z} + \textbf{a}$. \cite{liu2009false} shows that if the attacker knows the power system matrix $\textbf{H}$, she/he can construct $\textbf{a} = \textbf{H} \textbf{c}$ ($\textbf{c}$ represents the estimation error) that can bypass the fault detection in state estimation, as shown by equation (\ref{eq:FDIAPri}), where $\textbf{$\hat{\textbf{x}}$}_{bad}$ and $\textbf{$\hat{\textbf{x}}$}$ denote the estimated $\textbf{x}$ using $\textbf{z}_{a}$ and $\textbf{z}$ respectively. The details of FDIA can be found in \cite{liu2009false}.

\begin{subequations}
\begin{align}
\left \| \textbf{z}_{a} - \textbf{H} \textbf{$\hat{\textbf{x}}$}_{bad} \right \| \;\; &= \;\; \left \| \textbf{z} + \textbf{a} - \textbf{H}(\textbf{$\hat{\textbf{x}}$} + \textbf{c}) \right \| \\
 \;\; &= \;\; \left \| \textbf{z} - \textbf{H}\textbf{$\hat{\textbf{x}}$} + (\textbf{a} - \textbf{H}\textbf{c}) \right \| \\
 \;\; &= \;\; \left \| \textbf{z} - \textbf{H}\textbf{$\hat{\textbf{x}}$} \right \| \leq \tau
\end{align}
\label{eq:FDIAPri}
\end{subequations}

The above equation can be further represented as:

\begin{subequations}
\begin{align}
\textbf{a} = \textbf{H}\textbf{c} \;\; &\Leftrightarrow\;\; \textbf{P}\textbf{a} = \textbf{P}\textbf{H}\textbf{c} \Leftrightarrow \textbf{P}\textbf{a} = \textbf{H}\textbf{c} \Leftrightarrow \textbf{P}\textbf{a} = \textbf{a} \\
\;\; &\Leftrightarrow\;\; \textbf{P}\textbf{a} - \textbf{a} = \textbf{0} \Leftrightarrow  (\textbf{P}-\textbf{I})\textbf{a} = \textbf{0} \\
\;\; &\Leftrightarrow\;\; \textbf{B}\textbf{a} = \textbf{0}
\end{align}
\label{eq:FDIAgen}
\end{subequations}

 where $\textbf{P} = \textbf{H}(\textbf{H}^{T}\textbf{H})^{-1}\textbf{H}^{T}$ and matrix $\textbf{B} = \textbf{P} - \textbf{I}$.

\subsection{Attack Properties}

In this paper, we mainly consider supervised learning techniques for FDIA detection as they are more popular in the literature \cite{ozay2015machine, yan2016detection, he2017real, deng2018false, james2018online, ayad2018detection, ashrafuzzaman2018detecting, niu2019dynamic}. We note that our method can be easily extended to unsupervised learning approaches, such as the autoencoder used in \cite{wang2020detection}, by designing corresponding loss functions.

Without loss of generality, we consider the DNN-based FDIA detection as a binary classification problem. The trained DNN $F: \textbf{Z} \rightarrow Y$ maps the input measurement vectors \textbf{Z} to their labels $Y$. An ideal DNN will map a legitimate measurement $\textbf{z}$ to legitimate label $Y_{0}$ and a false measurements $\textbf{$\textbf{z}$}_{a} = \textbf{z} + \textbf{a}$ to false label $Y_{1}$.

\begin{table}[htbp]
\caption{Terminologies and Notations}
\begin{center}
\begin{tabular}{|c|c|c|}
\hline
\textbf{Terminology} & \textbf{Notation} & \textbf{Explanation}  \\
\hline
legitimate measurements & $\textbf{z}$ & sensors' original measurements  \\
\hline
injected false data & $\textbf{a}$ & meets constraint $\textbf{B}\textbf{a} = \textbf{0}$ \\
\hline
false measurements & $\textbf{$\textbf{z}$}_{a}$ & $\textbf{$\textbf{z}$}_{a}$ = $\textbf{z}$ + $\textbf{a}$\\
\hline
adversarial perturbation  & $\textbf{v}$ & perturbation for $\textbf{$\textbf{z}$}_{a}$ \\
\hline 
total injected false data & $\textbf{$\hat{\textbf{a}}$}$ & $\textbf{$\hat{\textbf{a}}$} = \textbf{a} + \textbf{v}$,  $\textbf{B}\textbf{$\hat{\textbf{a}}$} = \textbf{0}$  \\
\hline
adversarial measurements & $\textbf{$\textbf{z}$}_{adv}$ & $\textbf{$\textbf{z}$}_{adv} = \textbf{$\textbf{z}$}_{a} + \textbf{v} = \textbf{z} + \textbf{$\hat{\textbf{a}}$}$  \\

\hline
\end{tabular}
\end{center}
\label{table:notations}
\end{table}

The adversarial attack in FDIA detection is a false-negative attack that deceive $F$ to classify the false measurements as legitimate. The attacker needs to generate an adversarial perturbation vector $\textbf{v}$ and add it to the false measurement vector $\textbf{$\textbf{z}$}_{a}$, so that the adversarial measurement $\textbf{$\textbf{z}$}_{adv} = \textbf{$\textbf{z}$}_{a} + \textbf{v} = \textbf{z} + \textbf{a} + \textbf{v} = \textbf{z} + \textbf{$\hat{\textbf{a}}$}$ can be classified as legitimate measurements (with $Y_{0}$ label) by $F$, where $\textbf{$\hat{\textbf{a}}$} = \textbf{a} + \textbf{v}$ is the total injected false data. Meanwhile, the attacker requires $\textbf{$\hat{\textbf{a}}$}$ also meet the constraint $\textbf{B}\textbf{$\hat{\textbf{a}}$} = \textbf{0}$ defined by (\ref{eq:FDIAgen}c) to avoid being removed by the state estimation residual-based detection system. Therefore, the adversarial measurements in FDIA detection can also be considered as special false measurements. To be clear, Table \ref{table:notations} summarizes the terminologies and notations.

\subsection{Defense Requirements \& Threat Model}

White-box adversarial attacks allow the attacker to have access to the target DNN, which are common attacks in previous literature and have been extensively studied since they help researchers to learn the weakness of DNNs more directly \cite{xu2020adversarial}. Robust against white-box adversarial attacks is a desired property that the DNNs should maintain \cite{tramer2017ensemble}, especially for critical infrastructure like power grids.

In this paper, we expect our defense mechanism to be resilient to white-box FDIA adversarial attacks. 

From the attacker's point of view, we summarize the threat model of the adversarial attacks in FDIA detection:

\begin{itemize}

\item The attacker can compromise $k$ $(k > m-n)$ measurements in the power system, which is inherited from the FDIA attacks' requirement\cite{liu2009false}. 

\item Also as described in \cite{liu2009false}, the attacker knows the $\textbf{H}$ matrix to launch FDIA.

\item The total injected false data $\textbf{$\hat{\textbf{a}}$}$ should always follow $\textbf{B}\textbf{$\hat{\textbf{a}}$} = \textbf{0}$ to bypass the built-in residual-based detection mechanism in state estimation.

\end{itemize}

\subsection{Adversarial Attack Algorithm} \label{advalg}

We employ the iterative projection framework in \cite{kurakin2016adversarial} and \cite{madry2017towards} since the perturbation $\textbf{v}$ needs to follow the constraints defined by the power system. However, instead of projecting to the $\varepsilon$-neighbor ball, 
the attacker needs to map the adversarial perturbation $\textbf{v}$ to the solution space of the homogeneous equation $\textbf{B}\textbf{$\textbf{v}$} = \textbf{0}$ so that the total injected false data $\textbf{B}\textbf{$\hat{\textbf{a}}$} = \textbf{B}(\textbf{$\textbf{a}$} + \textbf{v}) = \textbf{B}\textbf{$\textbf{a}$} + \textbf{B}\textbf{v} = \textbf{0}$ will enable the adversarial measurements $\textbf{$\textbf{z}$}_{adv}$ to bypass the residual-based detection.

\section{Adversarial Defense in FDIA Detection} \label{sec:defense}

The arms race between adversarial attacks and defense in the AI domain is still in progress. In recent years, different adversarial defense mechanisms are proposed in the literature, such as model distillation \cite{papernot2016distillation}, adversarial training \cite{kurakin2016adversarial, shafahi2019adversarial, tramer2017ensemble}, adversarial detection \cite{ma2019nic, metzen2017detecting, xu2017feature}, and input reconstruction \cite{gu2014towards}\cite{meng2017magnet}. However, to the best of our knowledge, no defense method was demonstrated to be effective against all adversarial attacks \cite{xu2020adversarial}\cite{athalye2018obfuscated}. In particular, adversarial examples will always exist since none of the trained DNNs are perfect. Therefore, adversarial defense in FDIA detection mainly aims to degrade the attack performance and increase the attack cost.

In this section, we review three typical adversarial defense mechanisms proposed in the literature. We analyze these approaches and show that they have inherent limitations to mitigate adversarial attacks in FDIA detection effectively. We verify our analysis with simulations in Section \ref{sec:implementation}. After that, we propose our random input padding framework to train adversarial-resilient DNNs for FDIA detection. The evaluation in Section \ref{sec:implementation} shows that our framework can significantly decrease the overall performance of adversarial attacks.

\subsection{Typical Adversarial Defense Methods} \label{sec:sotadefense}

\subsubsection{\textbf{Defensive Distillation}}

Neural network distillation was originally proposed to reduce the size of DNN architectures \cite{hinton2015distilling}. In 2016, Papernot \etal employed distillation as an adversarial defense approach in the computer vision field \cite{papernot2016distillation}. The basic idea behind DNN distillation is to transfer the knowledge from a trained model to a new model, and the new DNN is shown to be less sensitive to input perturbations and then becomes more robust to adversarial attacks.

\noindent \textit{\textbf{Limitation Analysis:}} A common premise of adversarial attacks in the computer vision domain is that the size of the perturbation needs to be relatively small so that the adversarial pictures will not be noticed by human eyes. The assumption is no longer feasible in DNN-based FDIA detection. 
For example, a reckless attacker who aims to inject considerable false data into the state estimation only needs to consider the size of the total injected false data $\textbf{$\hat{\textbf{a}}$}$. Although model distillation decreases the DNN's sensitivity to input perturbations, the attacker can still iteratively search for a valid adversarial measurement $\textbf{z}_{adv}$ that deceives the DNN model without considering the size of $\textbf{v}$. Therefore, model distillation is not capable to defend against adversarial attacks in FDIA detection.

\subsubsection{\textbf{Adversarial Training}} \label{sec:advtrainingintro}

Adversarial training is one of the common methods to mitigate an adversarial attack \cite{kurakin2016adversarial}\cite{shafahi2019adversarial}\cite{tramer2017ensemble}. The basic principle of adversarial training is to generate and include adversarial examples in each data batch during the training stages. As the DNN is trained to recognize adversarial examples, it becomes more robust. 

\noindent \textit{\textbf{Limitation Analysis:}} Adversarial training needs to generate adversarial examples for each batch of data during the training process, which increases the training computation overhead significantly. As demonstrated in Section \ref{advalg}, to avoid being removed by the residual-based detection scheme, the adversarial perturbations must be projected to fit the constraint. The mapping process will further significantly introduce computation overhead to the adversarial training process. Therefore, adversarial training is not scalable to large systems that contain massive data resources. Meanwhile, \cite{kurakin2016adversarial} shows that adversarial training performs deficiently for iterative attacks, which makes it inappropriate for FDIA detection.

\subsubsection{\textbf{Adversarial Detection}} \label{sec:advDetectionAnalyze}

Adversarial detection recognizes adversarial examples at the DNN inference stage \cite{ma2019nic}\cite{metzen2017detecting}\cite{xu2017feature}. In particular, an auxiliary binary classification DNN $F_{adv}$ is trained with normal records and corresponding adversarial examples \cite{metzen2017detecting} to detect if an input is an adversarial example. The adversarial detection DNN $F_{adv}$ will be employed first to recognize the input records, and only the normal records will be fed into the original functional DNN. 

\noindent \textit{\textbf{Limitation Analysis:}} Adversarial detection assumes that the adversarial examples follow a different distribution from normal inputs. The assumption is reasonable in the computer vision domain (the natural images will not contain the well-crafted perturbations) but not applicable for FDIA detection. As introduced in Section \ref{sec:attack}, the manifold of the injected false data $\textbf{a}$ can be represented by the constraint $\textbf{B}\textbf{a} = \textbf{0}$ empirically. To bypass the built-in residual-based detection of state estimation, the total injected false data $\textbf{$\hat{\textbf{a}}$}$ is also required to meet the constraint $\textbf{B}\textbf{$\hat{\textbf{a}}$} = \textbf{0}$. Since the number of the possible attack scenarios (different meters can be compromised) can be large, intuitively, the crafted adversarial measurement vector $\textbf{z}_{adv}$ shares a similar manifold with the false measurement vector $\textbf{z}_{a}$ in the FDIA detection. In fact, $\textbf{z}_{adv}$ can be considered as special $\textbf{z}_{a}$, as analyzed in Section 4.1. Therefore,  adversarial detection will not work effectively in FDIA detection tasks. This analysis can also be adapted to input reconstruction methods \cite{gu2014towards}\cite{meng2017magnet}. 

\subsection{Random Input Padding Framework}

As discussed above, the adversarial defense in DNN-based FDIA detection is non-trivial since the adversarial measurements share the same manifold as the general false measurements. Given the victim model, the attacker generates the perturbation $\textbf{v}$ for $\textbf{z}_{a}$ iteratively through a gradient-based optimization process. As presented in \cite{kurakin2016adversarial}, the perturbation generated by multi-step attacks usually has worse transferability, which indicates that adversarial perturbation $\textbf{v}$ in FDIA detection is highly likely to be unique for each given $\textbf{z}_{a}$. Therefore, there is an intuition that the perturbation will no longer work if the input to the model changes.  Inspired by the stochastic-based defense mechanisms in the computer vision field \cite{meng2017magnet}\cite{xie2018mitigating}, we propose a random input padding defense framework to mitigate the effect of adversarial attacks in FDIA detection.

\begin{figure}[htbp]
\centerline{\includegraphics[width=0.95\linewidth]{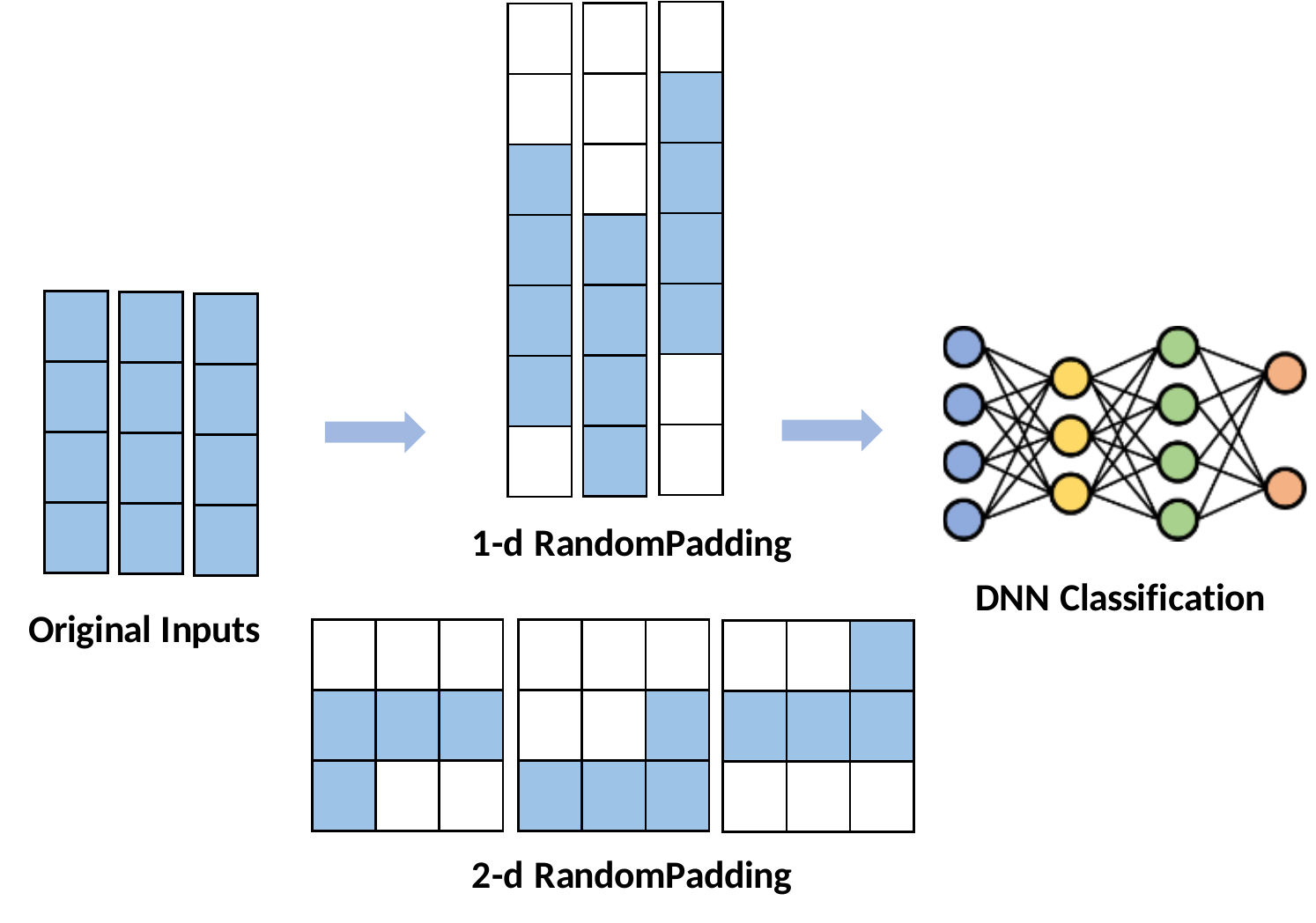}}
\caption{Illustration of random inputs padding framework. Each record in the plain inputs is a 4-d vector. The above 1-d padding scheme randomly pads four zeros to plain inputs and the padded vectors become an 8-d $(P = 8)$. The below 2-d padding scheme reshapes the inputs pads five zeros, which finally generates $3 \times 3$ dimensional records $(P = 9)$.}
\label{fig:randomPad}
\end{figure}

The philosophy of our random input padding framework is straightforward, and the overall structure is shown in Fig \ref{fig:randomPad}. A random padding layer is added in front of the DNN in both training and inference stages. In general, the measurements of the sensors $\textbf{z}$ are used as the features to train the detection models. Our framework firstly requires the operator to pick a padding dimension number $P$ $(P > m)$ as the input feature numbers for the DNN. Thereafter, we pad $P - m$ zeros randomly to the plain inputs $\textbf{z}$ and there will be $P - m$ padding scenarios in total. The DNN is then required to learn the pattern from the plain measurements that are embedded into the padded inputs during the training process. During the inference stage, when a new measurement vector $\textbf{z}$ is received, the framework randomly pads $\textbf{z}$ to a $P$ dimensional vector and feeds the padded vector to the DNN. Ideally, the detection rate against adversarial attacks should be $1 - \frac{1}{P - m + 1}$ $(P \geq m)$. The padding framework also works with possible input reshape, as shown in Fig \ref{fig:randomPad}.

As the padding process is random for each $\textbf{z}$ at the inference stage, the attacker (and even the operator) cannot know the final DNN padded input vectors even when she/he knows the whole framework. The attacker will be able to generate perturbations for one of the $P - m$ padding scenarios. Since the multi-steps perturbations have relatively weak transferability, the adversarial attacks should have a lower success rate under the random padding framework. Intuitively, a larger $P$ will decrease the success rate of adversarial attacks and finally increase the robustness of the DNN used for FDIA detection.

Different from \cite{xie2018mitigating}, our framework requires input data pre-processing (padding) during the training stage and cannot be applied to a trained model directly. This is because the measurement data of a specific power system should follow the manifold defined by the physical property of the system, which will be destroyed if the measurement vectors are reshaped, resized, or sampled directly.  On the other hand, the FDIA detection performance of the legitimate data vector $\textbf{z}$ and false measurements $\textbf{z}_{a}$ will not be constrained by the padding/scale size if an appropriate structure of the neural network is selected. Meanwhile, our framework only increases the computation of the training process slightly and is compatible with different neural networks.

\section{Simulation} \label{sec:implementation}

In this section, we first evaluate the effect of adversarial attacks in DNN-based FDIA detection. After that, we evaluate the typical adversarial defense approaches discussed in Section \ref{sec:sotadefense} and verify our analysis of their limitation through numeric simulations. Finally, we evaluate the performance of our random padding framework.

\noindent \textbf{Synthetic Dataset Generation:} The simulations are conducted based on the standard IEEE 118-bus system, which was used as a test benchmark in previous literature \cite{liu2009false, ozay2015machine, he2017real, james2018online}. We employ the MATPOWER \cite{zimmerman2010matpower} tool to derive the $\textbf{H}$ matrix of the system and simulate the power flow measurement data of each branch as $\textbf{Z}$. In our evaluation, each measurement vector $\textbf{z}$ contains $m = 186$ measurements. The FDIA is also implemented with MATLAB based on the corresponding $\textbf{H}$. We simulate a dataset $D_{train}$ that contains 30,000 legitimate measurement vectors. We pollute half of the records in $D_{train}$ by injecting false data generated by the FDIA. To simulate practical possible attacks, we set different compromised numbers $70<k<100$, and the indexes of the compromised measurements are randomly selected. In addition to $D_{train}$, we generate four test datasets. We consider the scenarios that there were 75, 80, 85, 90 measurements being compromised by the attacker respectively, and simulate 1000 polluted (false) data for each scenario.

\noindent \textbf{Target DNN:} We train a feed-forward neural network $F$ as the target DNN in our simulations, and the structure of $F$ is shown in Table \ref{table:FFW}. We randomly split $D_{train}$ into the training part and testing part, with the test part containing 15\% of records in $D_{train}$. We employ the categorical cross-entropy as the loss function and utilize stochastic gradient descent to optimize the loss. Through tuning parameters, $F$ finally achieves a 98.9\% detection accuracy and 98.6\% recall. $F$ is developed with the Tensorflow and Keras libraries. The simulations are conducted on a Windows 10 machine with an Intel i7 CPU and an additional NVIDIA GeForce GTX 1070 GPU to accelerate the training process.

\begin{table}[htbp]
\caption{Model Structure of $F$}
\begin{center}
\begin{tabular}{c|c|c|c|c|c|c}
\toprule[2pt]
\textbf{Layer} & \textbf{1} & \textbf{2} & \textbf{3} & \textbf{4} & \textbf{5} & \textbf{6} \\
\hline
\textbf{Nodes} & 186 & 128 & 64 & 16 & 0.25 Dropout & 2 Softmax \\
\bottomrule[2pt]
\multicolumn{7}{l}{$\ast$ Dense layer is used for each layer.} \\
\multicolumn{7}{l}{$\ast$ The activation function is $ReLu$ unless specifically noted.} \\
\end{tabular}
\end{center}
\label{table:FFW}
\end{table}

\noindent \textbf{Evaluation Metrics:} We set three metrics to evaluate the attack performance. The first metric is the detection \textbf{Recall} of the target DNN under adversarial attacks, which represents the probability of the adversarial measurements fooling the target DNN. In addition, we consider two different attack scenarios. The attack scenario happens when the attacker aims to inject a specific false vector $\textbf{a}$ into state estimation, such as to gain specific profit through modifying local marginal price \cite{bi2013false}\cite{xie2010false}. In this scenario, the perturbation $\textbf{v}$ is required to be small and our second metric is the $L_2$-Norm of $\textbf{v}$ of effective adversarial measurements that successfully fool the DNN's detection. We denote this metric as \textbf{Bias} \textbf{$L_2$-Norm}. The second scenario is that a malicious attacker aims to inject considerable false data to state estimation and the size of $\textbf{$\hat{\textbf{a}}$}$ is expected to be large. We set the $L_2$-Norm of $\textbf{$\hat{\textbf{a}}$}$ to be the third evaluation metric, which is noted as \textbf{Valid} \textbf{$L_2$-Norm}. In summary, from the attacker's point of view, a lower Recall, a smaller Bias $L_2$-Norm, and a larger Valid $L_2$-Norm indicate a more successful adversarial attack and vice versa.

\subsection{Adversarial Attacks in FDIA Detection} \label{sec:advAttackImp}

This subsection studies the vulnerabilities of DNN through adversarial attacks. As described above, the size of the simulated FDIA noise in $D_{train}$ follows a Gaussian distribution. Therefore, it is intuitive that a larger injected false data $\textbf{a}$ will result in a higher probability to be detected by $F$. On the contrary, the false measurements $\textbf{z}_a$ become difficult to be distinguished by $F$ if $\textbf{a}$ is small, but the FDIA performance also becomes worse.

\begin{figure}[htbp]
\centerline{\includegraphics[width=0.85\linewidth]{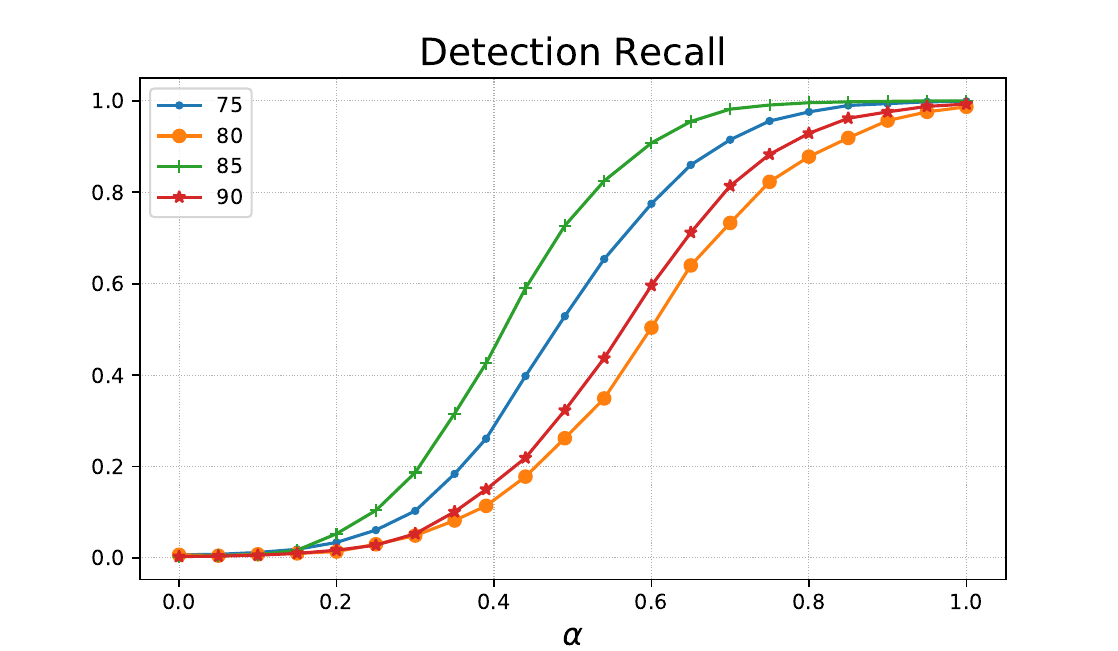}}
\caption{Vanilla attack recall.}
\label{fig:vanillaRecall}
\end{figure}

\begin{figure}[htbp]
\centerline{\includegraphics[width=0.99\linewidth]{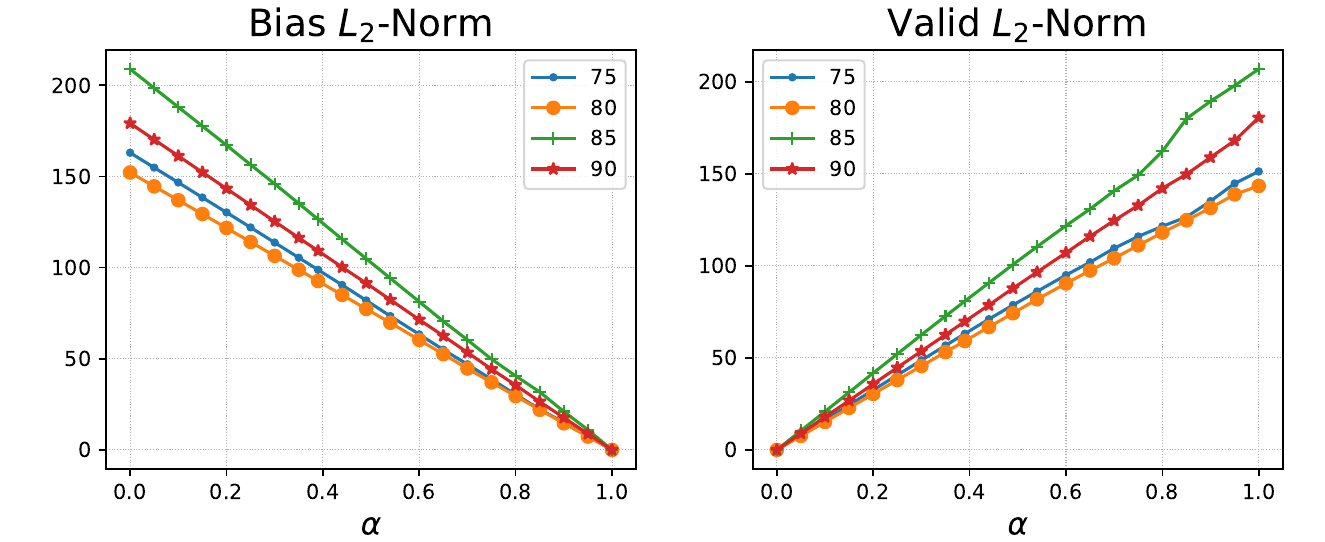}}
\caption{Vanilla attack bias.}
\label{fig:vanillaNoise}
\end{figure}

\noindent \textbf{Baseline:} To demonstrate the effectiveness of adversarial attacks, we set a vanilla attack as a baseline. The vanilla attack will simply multiply the injected false data $\textbf{a}$ with a factor $\alpha$ in the four test datasets, and the newly crafted false measurements are fed into $F$ for evaluation. The evaluation results of vanilla attack are shown in Fig. \ref{fig:vanillaRecall} and Fig. \ref{fig:vanillaNoise} respectively. As analyzed, from Fig. \ref{fig:vanillaRecall} we can learn that a larger $\alpha$ will result in a higher detection recall of the false measurements, and the detection performance may be various for different attack scenarios. Fig. \ref{fig:vanillaNoise} demonstrates that the trend of Bias $L_2$-Norm and Valid $L_2$-Norm of the vanilla attack is adverse with the $\alpha$ increases under vanilla attack. Overall, if the vanilla attacker aims to obtain a relatively higher probability to bypass the $F$ detection, she/he obtains a high Bias $L_2$-Norm and low Valid $L_2$-Norm.

We then evaluate the performance of adversarial attacks, as summarized in Table \ref{table:plainEva}. The evaluation demonstrates that the adversarial attacks can decrease the detection recall significantly with a relatively low bias $L_2$-Norm and high valid $L_2$-Norm. The attack performance can be various for different attack scenarios and slightly affected by the attack parameters \textbf{step size} ($size$) \cite{kurakin2016adversarial, madry2017towards, li2021conaml}. By comparing Table \ref{table:plainEva} with Fig. \ref{fig:vanillaRecall} and Fig. \ref{fig:vanillaNoise}, we can learn that the adversarial attacks significantly out-performs vanilla attacks.

\begin{table}[htbp]
\caption{Attack Performance of Plain DNN}
\begin{center}
\begin{tabular}{c|c|c|c|c}
\toprule[2pt]
\textbf{Case} & \textbf{Size} & \textbf{Recall} &  \textbf{Bias $L_{2}$-Norm} & \textbf{Valid $L_{2}$-Norm}\\
\midrule[1pt]

\multirow{3}*{75} & 0.1 & 3.9\% & 79.7  & 112.2 \\
\cline{2-5}
  ~  & 0.5 & 0.9\% & 87.2 & 111.29 \\
\cline{2-5}
 ~ & 1.0 & 1.9\% & 89.6 & 108.9 \\
 \hline
 
\multirow{3}*{80} & 0.1 & 18.9\% & 64.5  & 114.4 \\
\cline{2-5}
  ~  & 0.5 & 12.9\% & 76.7 & 117.5 \\
\cline{2-5}
 ~ & 1.0 & 10.9\% & 80.2 & 114.1 \\
 \hline
 
 \multirow{3}*{85} & 0.1 & 7.0\% & 115.2  & 134.4 \\
\cline{2-5}
  ~  & 0.5 & 3.9\% & 124.0 & 134.3 \\
\cline{2-5}
 ~ & 1.0 & 3.9\% & 127.0 & 131.4 \\
 \hline
 
 \multirow{3}*{90} & 0.1 & 12.9\% & 79.7  & 165.5 \\
\cline{2-5}
  ~  & 0.5 & 3.9\% & 109.9 & 178.0 \\
\cline{2-5}
 ~ & 1.0 & 5.9\% & 109.9 & 178.0 \\
\bottomrule[2pt]

\end{tabular}
\end{center}
\label{table:plainEva}
\end{table}

\subsection{Typical Adversarial Defense Methods}

\subsubsection{Defensive Distillation}

\begin{figure}[htbp]
\centerline{\includegraphics[width=0.99\linewidth]{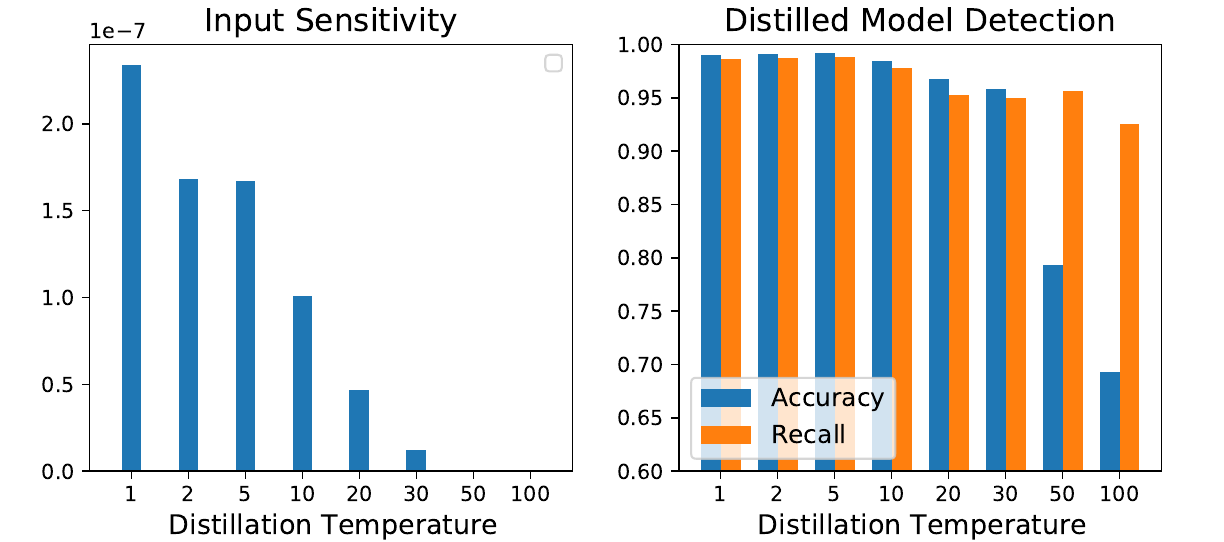}}
\caption{Distilled DNNs performance.}
\label{fig:distill-model}
\end{figure}

Similar to \cite{papernot2016distillation}, we select different distillation temperatures (1, 2, 5, 10, 20, 30, 50, 100) to train the corresponding distilled DNN models. Fig. \ref{fig:distill-model} demonstrates the properties of the distilled models. The left figure in Fig. \ref{fig:distill-model} shows that the input sensitivity decreases when the temperature increase, which coordinates with the effect of distillation, and the DNN models become less sensitive to input perturbation. From the right figure, we can learn that there is a slight decrease in both detection accuracy and recall of the distilled models with the temperature increases. However, when the temperature becomes larger, such as larger than 50, the detection accuracy decreases significantly, which indicates a high false-positive rate.

We evaluate the performance of distilled DNNs under adversarial attacks, as shown in Table \ref{table:dissEva}. The detection recall was low in most attack cases and the bias $L_2$-Norm and valid $L_2$-norm are comparable with plain DNN. Overall, model distillation presents ineffective defense performance to adversarial attacks in FDIA detection. 

\begin{table}[htbp]
\caption{Attack Performance of Distilled DNNs}
\begin{center}
\begin{tabular}{c|c|c|c|c}
\toprule[2pt]
\textbf{Temp} & \textbf{Case} & \textbf{Recall} &  \textbf{Bias $L_{2}$-Norm} & \textbf{Valid $L_{2}$-Norm}\\
\midrule[1pt]

\multirow{4}*{1} & 75 & 2.9\% & 77.2  & 116.8 \\
\cline{2-5}
  ~  & 80 & 57.9\% & 209.1 & 212.5 \\
\cline{2-5}
 ~ & 85 & 49.0\% & 207.0 & 248.7 \\
 \cline{2-5}
 ~ & 90 & 23.9\% & 149.1 & 225.1 \\
 \hline
 
 \multirow{4}*{2} & 75 & 3.9\% & 154.2  & 185.1 \\
\cline{2-5}
  ~  & 80 & 27.0\% & 104.2 & 185.5 \\
\cline{2-5}
 ~ & 85 & 34.9\% & 186.4 & 242.5 \\
 \cline{2-5}
 ~ & 90 & 3.9\% & 94.9 & 156.1 \\
 \hline

 \multirow{4}*{5} & 75 & 1.0\% & 90.5  & 108.6 \\
\cline{2-5}
  ~  & 80 & 5.9\% & 84.6 & 126.1 \\
\cline{2-5}
 ~ & 85 & 0.41\% & 195.8 & 148.9 \\
 \cline{2-5}
 ~ & 90 & 27.0\% & 158.6 & 252.1 \\
 \hline

 \multirow{4}*{10} & 75 & 5.9\% & 97.2  & 140.1 \\
\cline{2-5}
  ~  & 80 & 10.9\% & 71.2 & 144.0 \\
\cline{2-5}
 ~ & 85 & 43.0\% & 241.0 & 295.8 \\
 \cline{2-5}
 ~ & 90 & 7.9\% & 96.4 & 190.8 \\
 \hline

 \multirow{4}*{20} & 75 & 2.0\% & 57.6  & 139.5 \\
\cline{2-5}
  ~  & 80 & 12.9\% & 91.57 & 155.6 \\
\cline{2-5}
 ~ & 85 & 20.0\% & 145.1 & 175.4 \\
 \cline{2-5}
 ~ & 90 & 34.9\% & 179.2 & 271.6 \\
 \hline
 
\multirow{4}*{30} & 75 & 15.9\% & 94.6  & 157.8 \\
\cline{2-5}
  ~  & 80 & 11.9\% & 73.8 & 157.4 \\
\cline{2-5}
 ~ & 85 & 47.9\% & 203.7 & 222.3\\
 \cline{2-5}
 ~ & 90 & 41.9\% & 196.2 & 213.4 \\
 \hline
 
 \multirow{4}*{50} & 75 & 38.9\% & 172.9  & 233.8 \\
\cline{2-5}
  ~  & 80 & 28.9\% & 123.2 & 193.2 \\
\cline{2-5}
 ~ & 85 & 43.0\% & 215.1 & 212.4 \\
 \cline{2-5}
 ~ & 90 & 40.0\% & 202.1 & 268.0 \\
 \hline
 
\multirow{4}*{100} & 75 & 20.0\% & 136.2  & 196.1 \\
\cline{2-5}
  ~  & 80 & 18.0\% & 102.1 & 181.6 \\
\cline{2-5}
 ~ & 85 & 36.0\% & 201.7 & 230.9 \\
 \cline{2-5}
 ~ & 90 & 10.9\% & 74.6 & 202.0 \\
 
\bottomrule[2pt]
 \multicolumn{5}{l}{$\ast$ Parameters: $size = 0.1$.}
\end{tabular}
\end{center}
\label{table:dissEva}
\end{table}

\subsubsection{Adversarial Training}

\begin{figure}[htbp]
\centerline{\includegraphics[width=0.95\linewidth]{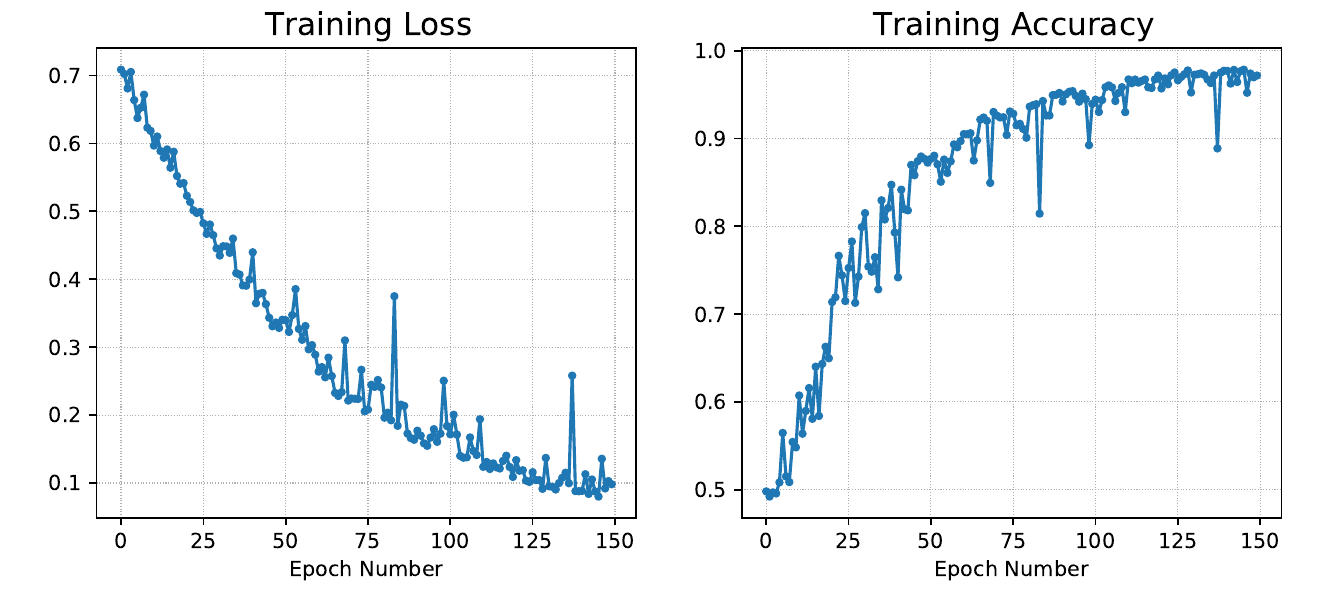}}
\caption{Adversarial training process.}
\label{fig:advModel}
\end{figure}

We follow the method described in \cite{kurakin2016adversarialphysical} to implement the adversary training. During each training epoch, we generate the adversarial measurements of the mini-batch measurement data with the real-time trained DNN and label them as false. We then add the generated adversarial measurements to the mini-batch data and train the model. Meanwhile, we project the generated measurements to follow the linear constraints defined by Equation (\ref{eq:FDIAgen}). Fig. \ref{fig:advModel} shows the training loss and detection accuracy during adversarial training. 

As analyzed in Section \ref{sec:advtrainingintro}, the projection process significantly increases the training computation overload, which makes it impractical to be employed in large-scale power systems. In our evaluation, the training process takes around 650 seconds to converge and achieves 96.5\% overall detection accuracy. For comparison, the normal training process takes around 10 seconds to converge in the same computer.

\begin{table}[htbp]
\caption{Attack Performance of Adversarial Training}
\begin{center}
\begin{tabular}{c|c|c|c|c}
\toprule[2pt]
\textbf{Case} & \textbf{Size} & \textbf{Recall} &  \textbf{Bias $L_{2}$-Norm} & \textbf{Valid $L_{2}$-Norm}\\
\midrule[1pt]

\multirow{3}*{75} & 0.1 & 15\% & 96.2  & 103.9 \\
\cline{2-5}
  ~  & 0.5 & 7.0\% & 115.8 & 113.8 \\
\cline{2-5}
 ~ & 1.0 & 5.9\% & 122.8 & 115.8 \\
 \hline
 
\multirow{3}*{80} & 0.1 & 43.1\% & 147.1  & 1.53 \\
\cline{2-5}
  ~  & 0.5 & 45.7\% & 359.9 & 354.4 \\
\cline{2-5}
 ~ & 1.0 & 47.2\% & 640.8 & 635.1 \\
 \hline
 
 \multirow{3}*{85} & 0.1 & 63.9\% & 142.0  & 167.5 \\
\cline{2-5}
  ~  & 0.5 & 61.0\% & 471.8 & 437.5 \\
\cline{2-5}
 ~ & 1.0 & 62.9\% & 519.0 & 678.3 \\
 \hline
 
 \multirow{3}*{90} & 0.1 & 62.2\% & 187.6  & 231.2 \\
\cline{2-5}
  ~  & 0.5 & 66.4\% & 761.8 & 768.4 \\
\cline{2-5}
 ~ & 1.0 & 69.3\% & 1485.9 & 1495.7 \\
\bottomrule[2pt]

\end{tabular}
\end{center}
\label{table:advTrainEva}
\end{table}

We evaluate the defense performance of adversarial training, we launched our adversarial attacks to the trained model, and the results are summarized in Table \ref{table:advTrainEva}. In our simulations, overall, adversarial training is demonstrated to increase the robustness of DNN to a certain degree. The detection recall of the adversarial trained DNN is higher than plain DNN in all evaluation scenarios. Meanwhile, although the Valid $L_2$-Norm values are very large in some scenarios (\textbf{Case 90}), the corresponding Bias $L_2$-Norm values are also large. However, for \textbf{Case 75}, the defense performance is still limited.

\subsubsection{Adversarial Detection}

\begin{figure}[htbp]
\centerline{\includegraphics[width=1\linewidth]{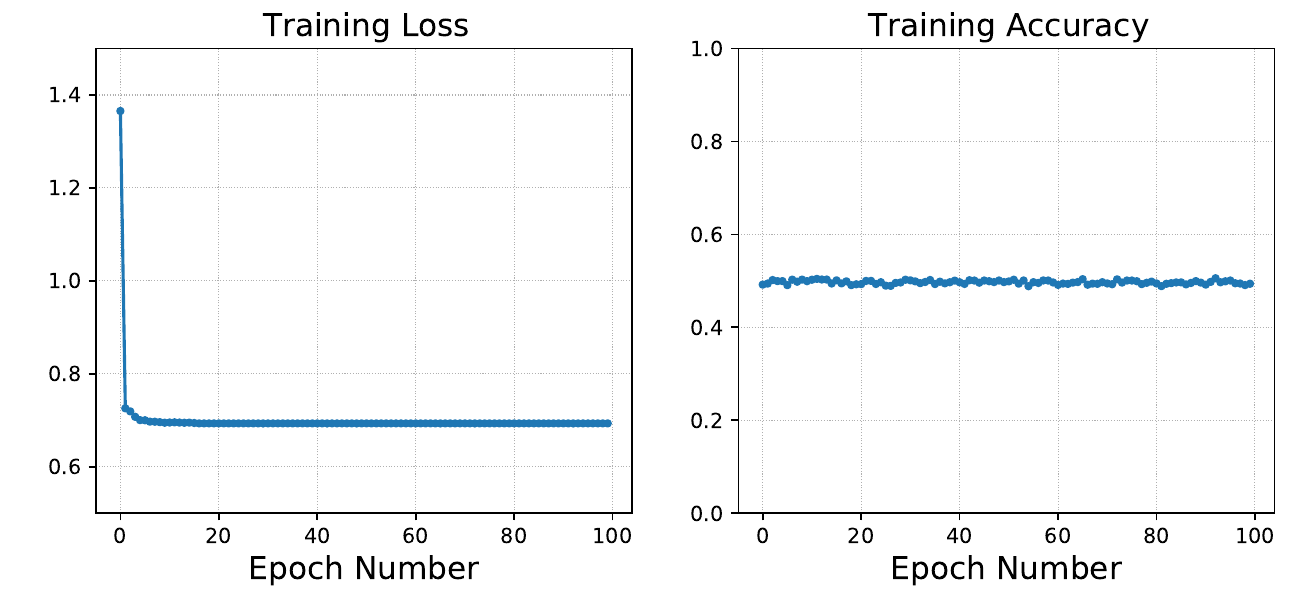}}
\caption{Training process of the auxiliary $f_{adv}$ for adversarial detection.}
\label{fig:advDetectTrain}
\end{figure}

We employ the adversarial detection methods described in \cite{metzen2017detecting} and generate the adversarial measurements of all the false measurements in $D_{train}$. We use the false measurements and their corresponding adversarial measurements to train a binary auxiliary classification DNN $F_{adv}$. We empirically attempt different structures and parameters of the $F_{adv}$ and observe that its performance is not reliable and the training process does not converge. As analyzed in Section \ref{sec:advDetectionAnalyze}, we explain that this result is caused by the similar manifolds shared between the false measurements $\textbf{z}_{a}$ and the adversarial measurements $\textbf{z}_{adv}$. Fig. \ref{fig:advDetectTrain} provides the training process of an example $F_{adv}$ in our simulation.

\begin{figure}[htbp]
\centerline{\includegraphics[width=1\linewidth]{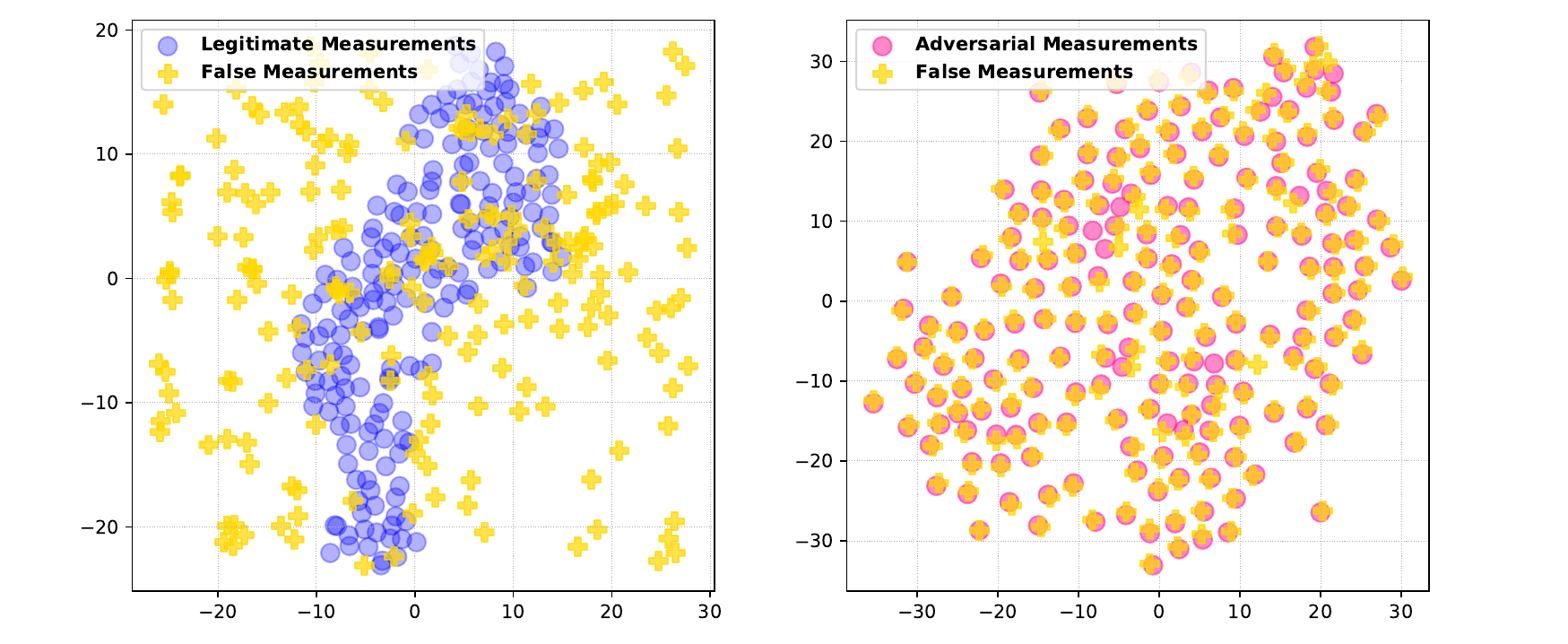}}
\caption{The manifolds of legitimate measurements, false measurements, and adversarial measurements with t-SNE dimensionality reduction.}
\label{fig:advDetectManifold}
\end{figure}

To verify our analysis, we utilize the t-Distributed Stochastic Neighbor Embedding (t-SNE) to visualize the manifolds in 2 dimensions, as shown in Fig. \ref{fig:advDetectManifold}. From the left figure, we can learn that the false measurements follow different manifolds with the legitimate measurements. This explains the effectiveness of DNN-based FDIA detection and the high detection recall of $F$. From the right figure, however, we can learn that the manifolds between the false measurements and corresponding adversarial measurements are very similar (overlapping markers). This phenomenon is caused by the physical property of the power system and the constraints defined by Equation (\ref{eq:FDIAgen}). The adversarial measurements $\textbf{z}_{adv}$ can be regarded as special false measurements. Therefore, adversarial detection can not distinguish the adversarial measurements from the DNN's inputs effectively in FDIA detection.

\subsection{Random Input Padding Framework}

\begin{figure}[htbp]
\centerline{\includegraphics[width=0.95\linewidth]{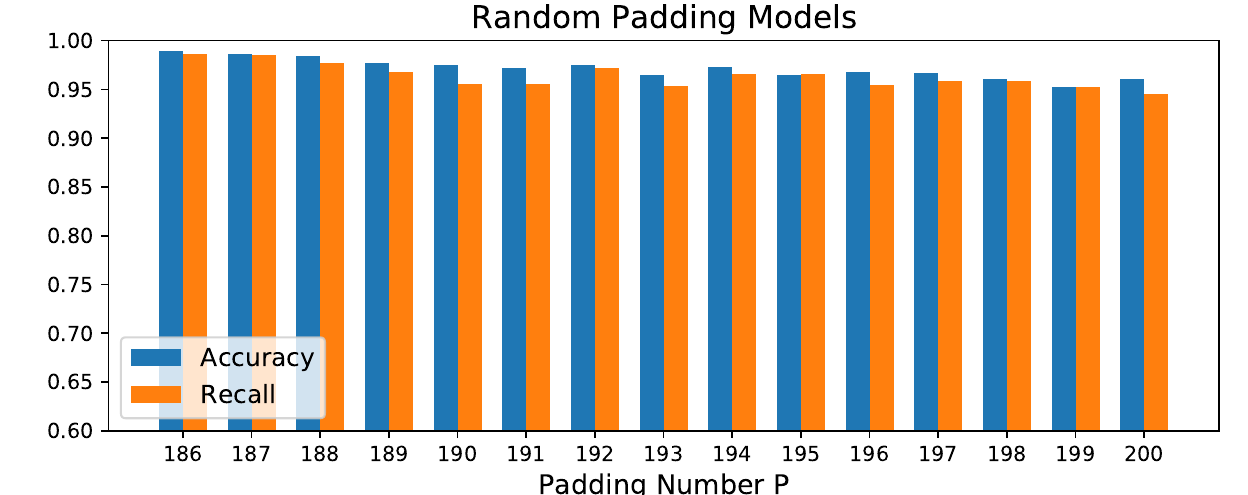}}
\caption{Detection performance of random input padding DNNs.}
\label{fig:padddingModel}
\end{figure}

\begin{figure*}[htbp]
\centerline{\includegraphics[width=0.95\linewidth]{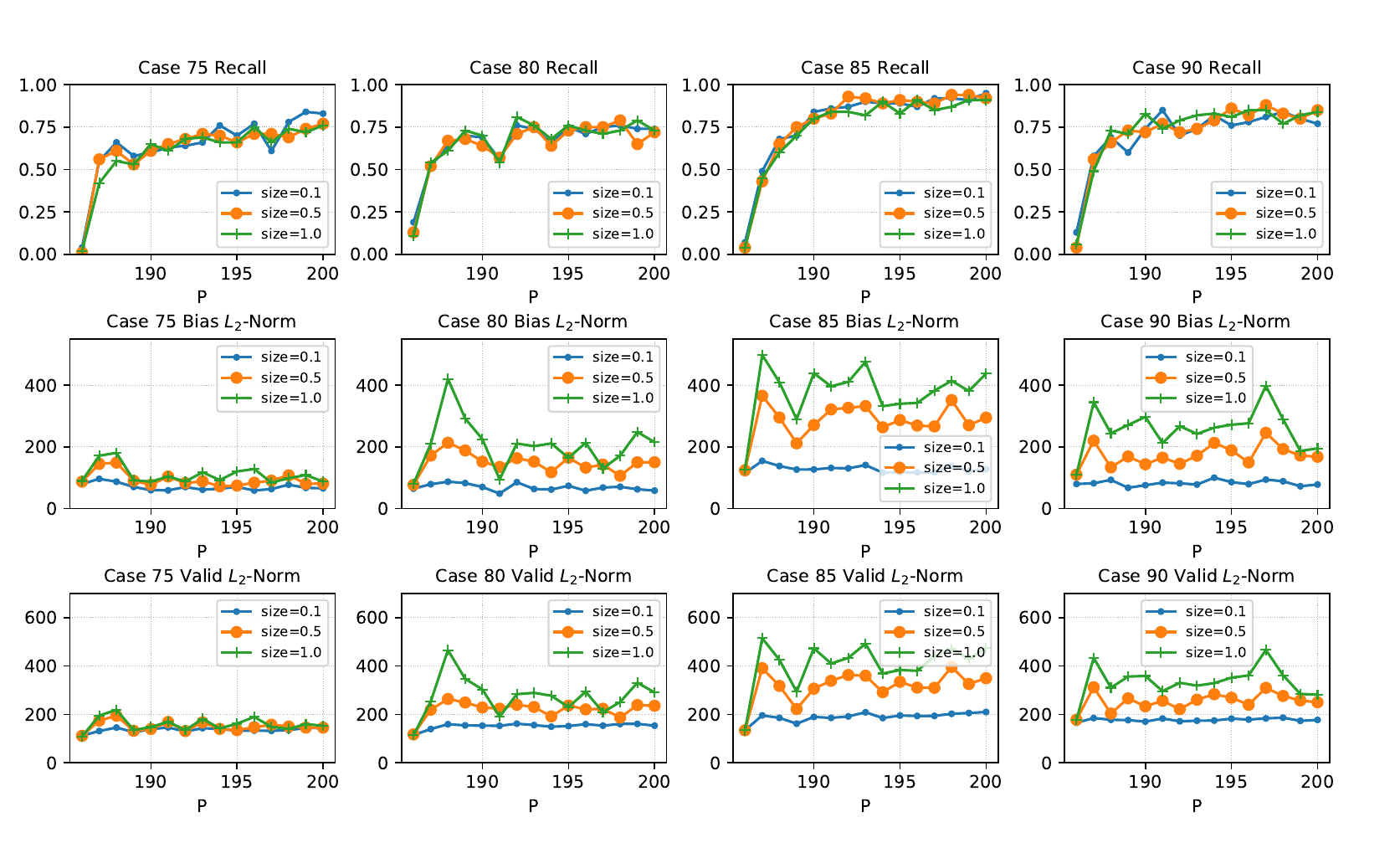}}
\caption{Adversarial evaluation of random input padding DNNs.}
\label{fig:padddingAttack}
\end{figure*}

We evaluate the defense performance of our framework based on $F$ and modify the number of the input neurons according to the padding number $P$ for $F$. The overall detection performance of padded DNNs are shown in Fig. \ref{fig:padddingModel}.  We can learn that with the padding number $P$ increases, the detection accuracy of DNN decreases gradually and becomes stable. The detection accuracy and recall of the padded DNNs reach around 96\% and 95\% respectively. Fig. \ref{fig:padddingModel} shows that the random padding framework only slightly decreases the FDIA detection performance, compared with the plain models ($P=186$). Our framework pads zeros in front of and/or after the plain measurements, which will not destroy the pattern of normal inputs.

Figure. \ref{fig:padddingAttack} shows the defense performance of the random input padding framework under different adversarial attacks. The top four figures in Fig. \ref{fig:padddingAttack} present the detection recall of the padded DNNs. The detection recalls of all attack scenarios increase remarkably when the padding number is relatively small ($P=187, 188$). After that, the detection recalls increase gradually and will converge to a specific range with the padding number increases. We can observe that the detection recalls under different attack scenarios do not strictly follow the expected $1 - \frac{1}{(P-186) + 1}$ $(P \geq 186)$ regulation. We note this is due to the transferability of adversarial measurements under different padding cases. Overall, we can observe that the padding framework significantly increases the resiliency of DNNs against adversarial attacks.

The bottom eight figures in Fig. \ref{fig:padddingAttack} demonstrate the bias $L_2$-Norm and valid $L_2$-Norm of the adversarial attacks. In general, we can observe that the bias $L_2$-Norm follows a similar trend with valid $L_2$-Norm. Compared with the plain DNN, our random padding framework can increase the bias $L_2$-Norm, which can decrease the performance of FDIA that targets local marginal price. On the contrary, similar to adversarial training, the valid $L_2$-Norm can also be large in some test scenarios. This phenomenon indicates that the attacker may inject considerable noise into the state estimation. However, our framework maintains a high detection recall under all test scenarios, which makes it outperform the other typical defensive methods discussed in this paper.

\section{Discussion and Future Work} \label{sec:discussion}

As we discussed in Section \ref{sec:defense}, currently, there is no defense mechanism that is robust to all adversarial attacks. In general, adversary examples/measurements will always exist since the DNN is imperfect and an attacker can always modify the input to force the DNN to change the prediction output. For example, if the vanilla attacker in Section \ref{sec:advAttackImp} sets the factor $\alpha$ to a small value, the resulted false measurements will have a high probability to bypass the DNN's detection. In this paper, we note the attacker can generate his/her adversarial perturbations based on multiple padding cases to increase their transferability and finally bypass the random padding system, as used in \cite{moosavi2017universal} and \cite{li2021conaml}. However, generating transferable perturbations will affect the attack performance, such as the valid $L_2$-Norm, and increase labor and resources of the attacker \cite{li2021conaml}. 

We proposed the input padding framework for FDIA detection in this paper and the framework is compatible with different models. Inspired by the randomly-selected autoencoder scheme in \cite{meng2017magnet}, we believe our framework can be used together with their system to further increase the DNNs' robustness in FDIA detection. In the future, we will study the implementation and performance of this joint system. 

In DNN-based FDIA detection, in addition to decreasing the detection recall, we expect the defense mechanism to increase the bias $L_2$-Norm and decrease the valid $L_2$-Norm. In our simulations, the random padding framework increases both the bias $L_2$-Norm and the valid $L_2$-Norm (similar to adversarial training). We consider this as an open problem in our future work and will study defense methods that achieve satisfying performance in all three metrics.

\section{Conclusion} \label{sec:conclusion}

The adversarial attacks present a serious threat to DNN-based FDIA detection. In this paper, we study the defense mechanism of adversarial attacks in FDIA and methods to increase the robustness of corresponding DNN models. We first analyze the unique properties of adversarial attack and defense in FDIA detection and summarize the defense requirements.  We evaluate several typical defense methods and demonstrate that they have intrinsic limitations for this task through analysis and numeric simulations. After that, we propose a random padding framework for employing DNNs in FDIA detection. The framework is compatible with different DNNs and introduces little extra computation. Evaluation results present that our framework outperforms the typical defense mechanisms and can significantly increase the robustness of DNNs against adversarial measurements in FDIA detection. Meanwhile, the framework will hardly decrease DNNs' detection performance on the normal inputs.

\section*{Acknowledgment}

This work was partially supported by the US National Science Foundation (NSF) under grant CNS-2038922. Meanwhile, this research was supported in part by the Engineering Research Center Program of the National Science Foundation and the Department of Energy under NSF Award Number EEC-1041877 and the CURENT Industry Partnership Program. We also thank Mr. Eric Reinsmidt and Mr. Jin Young Lee for their suggestions for improving the quality of this paper.

\bibliography{reference} 
\bibliographystyle{ieeetr}

\end{document}